\documentclass[prb,10pt]{revtex4}
\usepackage{amsfonts}
\usepackage[T1]{fontenc}
\usepackage{amsmath,amsbsy,amssymb,graphicx}
\usepackage{times}

\let\mathbf=\boldsymbol
\def\Section#1{\bigskip\noindent{\usefont{T1}{phv}{b}{n}\large#1}\par}

\begin{document}

\begin{flushleft}
{\textbf{{\Large Second-order topological insulators and loop-nodal semimetals
in Transition Metal Dichalcogenides XTe$_{2}$ (X=Mo,W)}}}
\end{flushleft}

\leftline{Motohiko Ezawa$^*$} 
\leftline{Department of Applied Physics, University of Tokyo, Hongo 7-3-1, 113-8656, Japan}
\leftline{$^*$Correspondence to ezawa@ap.t.u-tokyo.ac.jp}

\bigskip \Section{Abstract} Transition metal dichalcogenides XTe$_{2}$
(X=Mo,W) have been shown to be second-order topological insulators based on
first-principles calculations, while topological hinge states have been
shown to emerge based on the associated tight-binding model. The model is
equivalent to the one constructed from a loop-nodal semimetal by adding mass
terms and spin-orbit interactions. We propose to study a chiral-symmetric
model obtained from the original Hamiltonian by simplifying it but keeping
almost identical band structures and topological hinge states. A merit is
that we are able to derive various analytic formulas because of chiral
symmetry, which enables us to reveal basic topological properties of
transition metal dichalcogenides. We find a linked loop structure where a
higher linking number (even 8) is realized. We construct
second-order topological semimetals and two-dimensional second-order
topological insulators based on this model. It is
interesting that topological phase transitions occur without gap closing
between a topological insulator, a topological crystalline insulator and a
second-order topological insulator. We propose to characterize them by
symmetry detectors discriminating whether the symmetry is preserved or not.
They differentiate topological phases although the
symmetry indicators yield identical values to them. We also show that
topological hinge states are controllable by the direction of magnetization.
When the magnetization points the $z$ direction, the hinges states shift,
while they are gapped when it points the in-plane direction. Accordingly,
the quantized conductance is switched by controlling the magnetization
direction. Our results will be a basis of future topological devices based
on transition metal dichalcogenides.

\Section{Introduction}

Higher-order topological insulators (HOTIs) are generalization of
topological insulators (TIs). In the second-order topological insulators
(SOTIs), for instance, topological corner states emerge though edge states
do not in two dimensions, while topological hinge states emerge though
surface states do not in three dimensions\cite%
{Fan,Science,APS,Peng,Lang,Lin,Song,Bena,Schin,FuRot,EzawaKagome,Gei,MagHOTI,Kha,HexaHOTI}%
. The emergence of these modes is protected by symmetries and topological
invariants of the bulk. Hence, an insulator so far considered to be trivial
due to the lack of the topological boundary states can actually be a HOTI.
Indeed, phosphorene is theoretically shown to be a two-dimensional (2D) SOTI%
\cite{EzawaPhos}. A three-dimensional (3D) SOTI is experimentally realized
in rhombohedral bismuth\cite{Bis}, where topological quantum chemistry is
used for the material prediction\cite{TQP}. Transition metal dichalcogenides
XTe$_{2}$ (X=Mo,W) are also theoretically shown to be 3D SOTIs\cite%
{MoTe,MoTe2}.

The tight-binding model for transition metal dichalcogenides has already
been proposed, which is closely related to a type of loop-nodal semimetals%
\cite{MoTe}. A loop-nodal semimetal is a semimetal whose Fermi surfaces form
loop nodes\cite{FangLoop,Kim,Yu,Chan,CFang}. Especially, the Hopf semimetal
is a kind of loop-nodal semimetal whose Fermi surfaces are linked and
characterized by a nontrivial Hopf number\cite%
{WChen,ZYan,PYChang,EzawaHopf,HasanPRL}. There is another type of loop
nodal-semimetals characterized by the monopole charge\cite{FangLoop}. An
intriguing feature is that loop nodes at the zero-energy and another energy
form a linked-loop structure. The proposed model\cite{MoTe} may be obtained
by adding certain mass terms to this type of loop-nodal semimetals.

It is intriguing that topological boundary states can be controllable
externally. Magnetization is an efficient way to do so. Famous examples are
surface states of 3D magnetic TIs\cite{MagTI1,MagTI2,MagTI3,MagTI4}, where
the gap opens for out-of-plane magnetization, while the Dirac cone shifts
for in-plane magnetization. Similar phenomena also occur in 2D TIs, which
can be used as a giant magnetic resistor\cite{Rachel}. Recently, a
topological switch between a SOTI and a topological crystalline insulator
(TCI) was proposed\cite{Switch}, where the emergence of topological corner
states is controlled by magnetization direction. We ask if a similar
magnetic control works in transition metal dichalcogenides.

In this paper, we investigate a chiral-symmetric limit of the original model%
\cite{MoTe} constructed in such a way that the simplified model has almost
identical band structures and topological hinge states as the original one.
Alternatively, we may consider that the original model is a small
perturbation of the chiral symmetric model. A great merit is that we are
able to derive various analytic formulas because of chiral symmetry, which
enable us to reveal basic topological properties of transition metal
dichalcogenides. We find that a linking structure with a higher
linking number is realized in the 3D model. We also study 2D SOTIs
and 3D second-order topological semimetals (SOTSMs) based on this model.
Depending on the way to introduce mass parameters there are three phases,
i.e., TIs, TCIs and SOTIs in the 2D model. We find that topological phase
transitions occur between these phases without band gap closing. Hence, the
transition cannot be described by the change of the symmetry indicators. We
propose symmetry detectors discriminating whether the symmetry is preserved
or not. They can differentiate these three topological phases. Furthermore,
we show that the topological hinge states in the SOTIs are controlled by
magnetization. When the magnetization direction is out of plane, the
topological hinge states only shift. On the other hand, when the
magnetization direction is in plane, the gap opens in the topological hinge
states.

\begin{figure}[t]
\centerline{\includegraphics[width=0.45\textwidth]{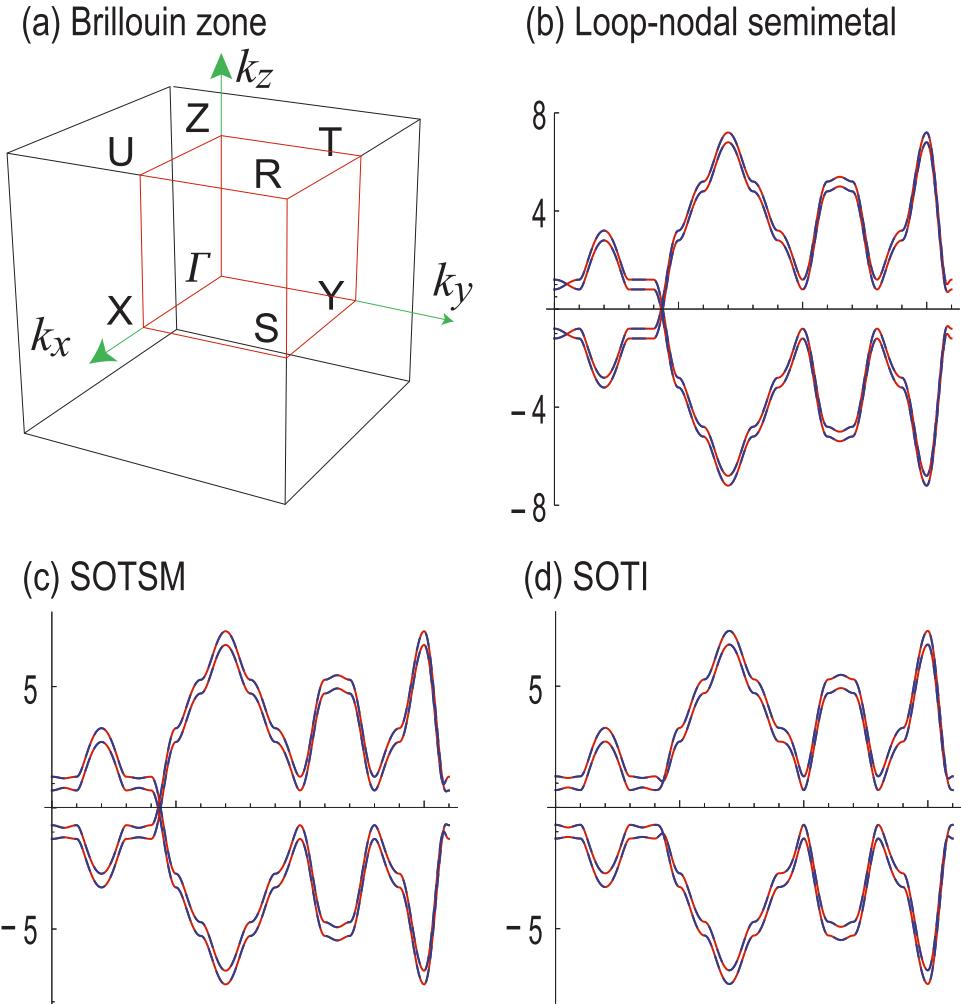}}
\caption{ (a) Brillouin zone and high symmetry points. (b)--(d) Bulk band
structures along the $\Gamma $-X-S-Y-$\Gamma $-Z-U-R-T-Z-Y-T-U-X-S-R-$\Gamma 
$ line (b) for loop-nodal semimetal, (c) for SOTSM and (d) for SOTI. There
are four bands in each phase. The dashed magenta curves represent the band
structure of the chiral-symmetric model, while the dashed cyan curves
represent that of the chiral-nonsymmetric model. They are indistinguishable
in these figures. We have chosen $t_{x}=t_{y}=1$, $t_{z}=2$, $\protect%
\lambda _{x}=\protect\lambda _{y}=1$, $\protect\lambda _{z}=1.2$, $m=-3$, $%
m_{2}=0.3$, $m_{3}=0.2$, $m_{mv1}=-0.4$, $m_{mv2}=0.2$, $m_{\text{Loop}}=0.3$
and $m_{\text{SOTSM}}=0.3$.}
\label{FigBrill}
\end{figure}

\Section{Result} 
\medskip\noindent\textbf{Hamiltonians.} 
Motivated by the model Hamiltonian\cite{MoTe} which describes the topological properties of
transition metal dichalcogenides $\beta $-(1T'-)MoTe$_{2}$ and $\gamma $%
-(Td-)XTe$_{2}$ (X=Mo,W), we propose to study a simplified model
Hamiltonian, 
\begin{equation}
H_{\text{SOTI}}=H_{0}+H_{\text{SO}}+V_{\text{Loop}}+V_{\text{SOTSM}},
\end{equation}
with 
\begin{eqnarray}
H_{0} &=&\left[ m+\sum_{i=x,y,z}t_{i}\cos k_{i}\right] \tau _{z}  \notag \\
&&+\lambda _{x}\sin k_{x}\tau _{x}+\lambda _{y}\sin k_{y}\tau _{y}\mu _{y},
\\
H_{\text{SO}} &=&\lambda _{z}\sin k_{z}\tau _{y}\mu _{z}\sigma _{z}, \\
V_{\text{Loop}} &=&m_{\text{Loop}}\tau _{z}\mu _{z},\quad V_{\text{SOTSM}%
}=m_{\text{SOTSM}}\mu _{x},
\end{eqnarray}
where $\sigma ,\tau $ and $\mu $ are Pauli matrices representing spin and
two orbital degrees of freedom. It contains three mass parameters, $m$, $m_{%
\text{Loop}}$ and $m_{\text{SOTSM}}$. The role of the term $m_{\text{Loop}}$
is to make the system a loop-nodal semimetal, and that of the term $m_{\text{%
SOTSM}}$ is to make the system a SOTSM. The Brillouin zone and high symmetry
points are shown in Fig.\ref{FigBrill}(a). Although the band structure of
the transition metal dichalcogenides is chiral nonsymmetric, the topological
nature is well described by the above simple tight-binding model.

\begin{figure*}[t]
\centerline{\includegraphics[width=0.45\textwidth]{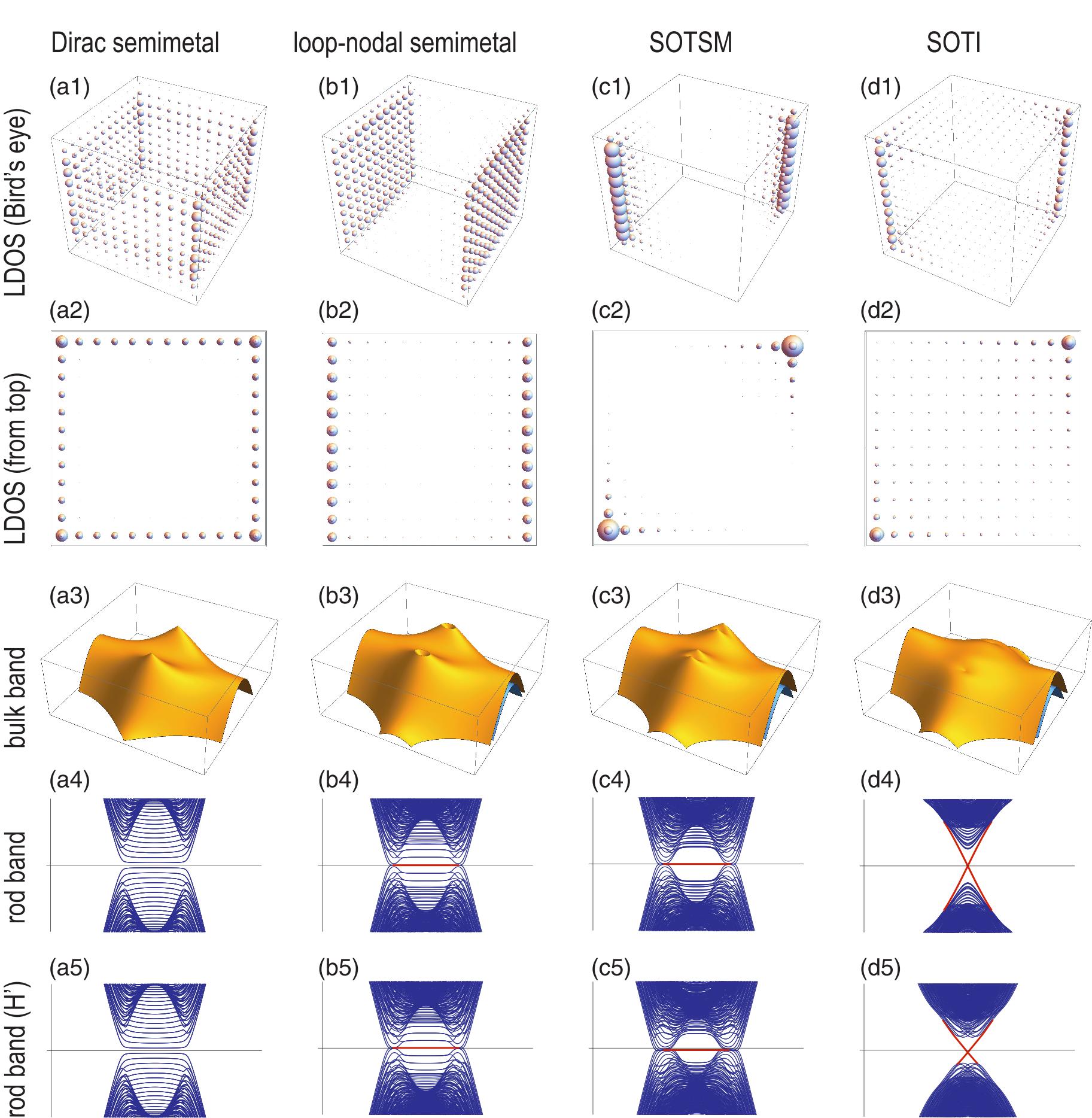}}
\caption{ (a1)--(d1) Bird's eye's views of the LDOS of the zero-energy
states: (a1) for $H_{0}$ with surface zero-energy states on the four side
surfaces; (b1) for $H_{\text{Loop}}$ with surface zero-energy states on the
two side surfaces; (c1) $H_{\text{SOTSM}}$ with hinge-arc states at two
pillars; (d1) $H_{\text{SOTI}}$ with hinge states at two pillars. (a2)--(d2)
Top view of the LDOS corresponding to (a1)--(d1). (a3)--(d3) Bulk band
structures of valence bands along $k_{x}=0$ plane for these Hamiltonians.
(a4)--(d4) Band structures of the square rod along $z$ direction for these
Hamiltonians. (a5)--(d5) Corresponding rod band structures for the
chiral-nonsymmetric Hamiltonian $H^{\prime }$. In these two sets of figures
red curves represent topological boundary states. The Parameters are the
same as in Fig.1.}
\label{FigCube}
\end{figure*}

The original Hamiltonian contains two extra mass parameters and given by 
\begin{equation}
H_{\text{SOTI}}^{\prime }=H_{0}+H_{\text{SO}}+V_{\text{Loop}}^{\prime }+V_{%
\text{SOTSM}}^{\prime }
\end{equation}
with 
\begin{eqnarray}
V_{\text{Loop}}^{\prime } &=&m_{2}\tau _{z}\mu _{x}+m_{3}\tau _{z}\mu _{z},
\\
V_{\text{SOTSM}}^{\prime } &=&m_{mv1}\mu _{z}+m_{mv2}\mu _{x}.
\end{eqnarray}
The simplified model $H_{\text{SOTI}}$ captures essential band structures of
the original model\ $H_{\text{SOTI}}^{\prime }$. Indeed, the bulk band
structures are almost identical, as seen in Fig.\ref{FigBrill}(b)--(d). The
rod band structures are also very similar, as seen in Fig.\ref{FigCube}%
(a4)--(d4) and (a5)--(d5), where the bulk band parts are found almost
identical while the boundary states (depicted in red) are slightly
different. Moreover, the both models have almost identical hinge states,
demonstrating that they describe SOTIs inherent to transition metal
dichalcogenides XTe$_{2}$.

A merit of the simplified model is the chiral symmetry, $\{H_{\text{SOTI}%
}(k_{x},k_{y},k_{z}),C\}=0$, which is absent in the original model, $\{H_{%
\text{SOTI}}^{\prime }(k_{x},k_{y},k_{z}),C\}\neq 0$. Accordingly, the band
structure of $H$ is symmetric with respect to the Fermi level. Moreover, the
bulk band structure is analytically solved. Here, the chiral symmetry
operator is $C=\tau _{y}\mu _{z}\sigma _{x}$ or $C=\tau _{y}\mu _{z}\sigma
_{y}$. Let us call the original model a chiral-nonsymmetric model and the
simplified model a chiral-symmetric model.

The common properties of the two Hamiltonians $H_{\text{SOTI}}$ and $H_{%
\text{SOTI}}^{\prime }$ read as follows. First, they have inversion symmetry 
$P=\tau _{z}$ and time-reversal symmetry $T=i\tau _{z}\sigma _{y}K$ with $K$
the complex conjugation operator. Inversion symmetry $P$ acts on $H_{\text{%
SOTI}}$ as $P^{-1}H_{\text{SOTI}}(\mathbf{k})P=H_{\text{SOTI}}(-\mathbf{k})$%
, while time-reversal symmetry $T$ acts as $T^{-1}H_{\text{SOTI}}(\mathbf{k}%
)T=H_{\text{SOTI}}(-\mathbf{k})$. Accordingly, the Hamiltonian has the $PT$
symmetry $\left( PT\right) ^{-1}H_{\text{SOTI}}(\mathbf{k})PT=H_{\text{SOTI}%
}(\mathbf{k})$, which implies that $H^*=H$. Second, the $z$-component of the
spin is a good quantum number $\sigma _{z}=s_{z}$. Since we may decompose
the Hamiltonian into two sectors, 
\begin{equation}
H_{\text{SOTI}}=H_{\text{SOTI}}^{\uparrow }\oplus H_{\text{SOTI}%
}^{\downarrow },  \label{SpinUD}
\end{equation}
it is enough to diagonalize the $4\times 4$ Hamiltonians. All these
relations hold also for $H_{\text{SOTI}}^{\prime }$. The relation (\ref%
{SpinUD}) resembles the one that the Kane-Mele model is decomposed into the
up-spin and down-spin Haldane models on the honeycomb lattice\cite%
{KaneMele,EzawaReview}.

A convenient way to reveal topological boundary states is to plot the local
density of states (LDOS) at zero energy. First, we show the LDOS for the
Hamiltonian $H_{0}$ in Fig.\ref{FigCube}(a1). It describes a Dirac
semimetal, whose topological surfaces appear on the four side surfaces.
Then, we show the LDOS for the Hamiltonian 
\begin{equation}
H_{\text{Loop}}=H_{0}+V_{\text{Loop}}
\end{equation}
in Fig.\ref{FigCube}(b1), where the topological surface states appear only
on the two side surfaces parallel to the $y$-$z$ plane. We will soon see
that a loop-nodal semimetal is realized in $H_{\text{Loop}}$. Next, we show
the LDOS for the Hamiltonian 
\begin{equation}
H_{\text{SOTSM}}=H_{0}+V_{\text{Loop}}+V_{\text{SOTSM}}
\end{equation}
in Fig.\ref{FigCube}(c1), where a SOTSM is realized with two topological
hinge-arcs. Finally, by including $H_{\text{SO}}$, we show the LDOS for the
Hamiltonian $H_{\text{SOTI}}$ in Fig.\ref{FigCube}(d1), where a SOTI is
realized with topological two-hinge state.

\medskip\noindent\textbf{Topological phase diagram.} 
The chiral-symmetric Hamiltonian $H_{\text{SOTI}}$ is analytically diagonalizable. The energy
dispersion is given by 
\begin{equation}
E=\pm \sqrt{F\pm \sqrt{G}}
\end{equation}
with 
\begin{eqnarray}
F &=&M^{2}+m_{\text{Loop}}^{2}+m_{\text{SOTSM}}^{2}  \notag \\
&&+\lambda _{x}^{2}\sin ^{2}k_{x}+\lambda _{y}^{2}\sin ^{2}k_{y}+\lambda
_{z}^{2}\sin ^{2}k_{z}, \\
G &=&\left( m_{\text{SOTSM}}\lambda _{x}\sin k_{x}-m_{\text{Loop}}\lambda
_{y}\sin k_{y}\right) ^{2} \\
&&+2M^{2}\left( m_{\text{Loop}}^{2}+m_{\text{SOTSM}}^{2}\right) ,
\end{eqnarray}
and 
\begin{equation}
M=m+\sum_{i=x,y,z}t_{i}\cos k_{i}.
\end{equation}

The topological phase diagram is determined by the energy spectra at the
eight high-symmetry points $\Gamma =\left( 0,0,0\right) $, $S=\left( \pi
,\pi ,0\right) $, $X=\left( \pi ,0,0\right) $, $Y=\left( 0,\pi ,0\right) $, $%
Z=\left( 0,0,\pi \right) $, $R=\left( \pi ,\pi ,\pi \right) $, $U=\left( \pi
,0,\pi \right) $ and $T=\left( 0,\pi ,\pi \right) $ with respect to
time-reversal inversion symmetry. The energies at these high-symmetry points 
$(k_{x},k_{y},k_{z})$ are analytically given by 
\begin{equation}
E\left( k_{i}\right) =\eta _{a}M\left( k_{i}\right) +\eta _{b}\sqrt{m_{\text{%
Loop}}^{2}+m_{\text{SOTSM}}^{2}},
\end{equation}%
where $\eta _{a}=\pm 1$ and $\eta _{b}=\pm 1$. The phase boundaries are
given by solving the zero-energy condition ($E=0$), 
\begin{equation}
\left( m+\eta _{x}t_{x}+\eta _{y}t_{y}+\eta _{z}t_{z}\right) ^{2}=m_{\text{%
Loop}}^{2}+m_{\text{SOTSM}}^{2},  \label{3DTopBound}
\end{equation}%
where $\eta _{x}=\pm 1$, $\eta _{y}=\pm 1$ and $\eta _{z}=\pm 1$. There are $%
16$ critical points apart from degeneracy. When $t_{x}=t_{y}$, the critical
points are reduced to be $12$ since $E\left( X\right) =E\left( Y\right) $
and $E\left( U\right) =E\left( T\right) $. Hence, solving $E=0$ for $t_{z}$,
there are $6$ solutions for $t_{z}>0$, which we set as $t_{n}$, $%
n=1,2,3,\cdots ,6$ with $t_{i}<t_{i+1}$.

\begin{figure}[t]
\centerline{\includegraphics[width=0.45\textwidth]{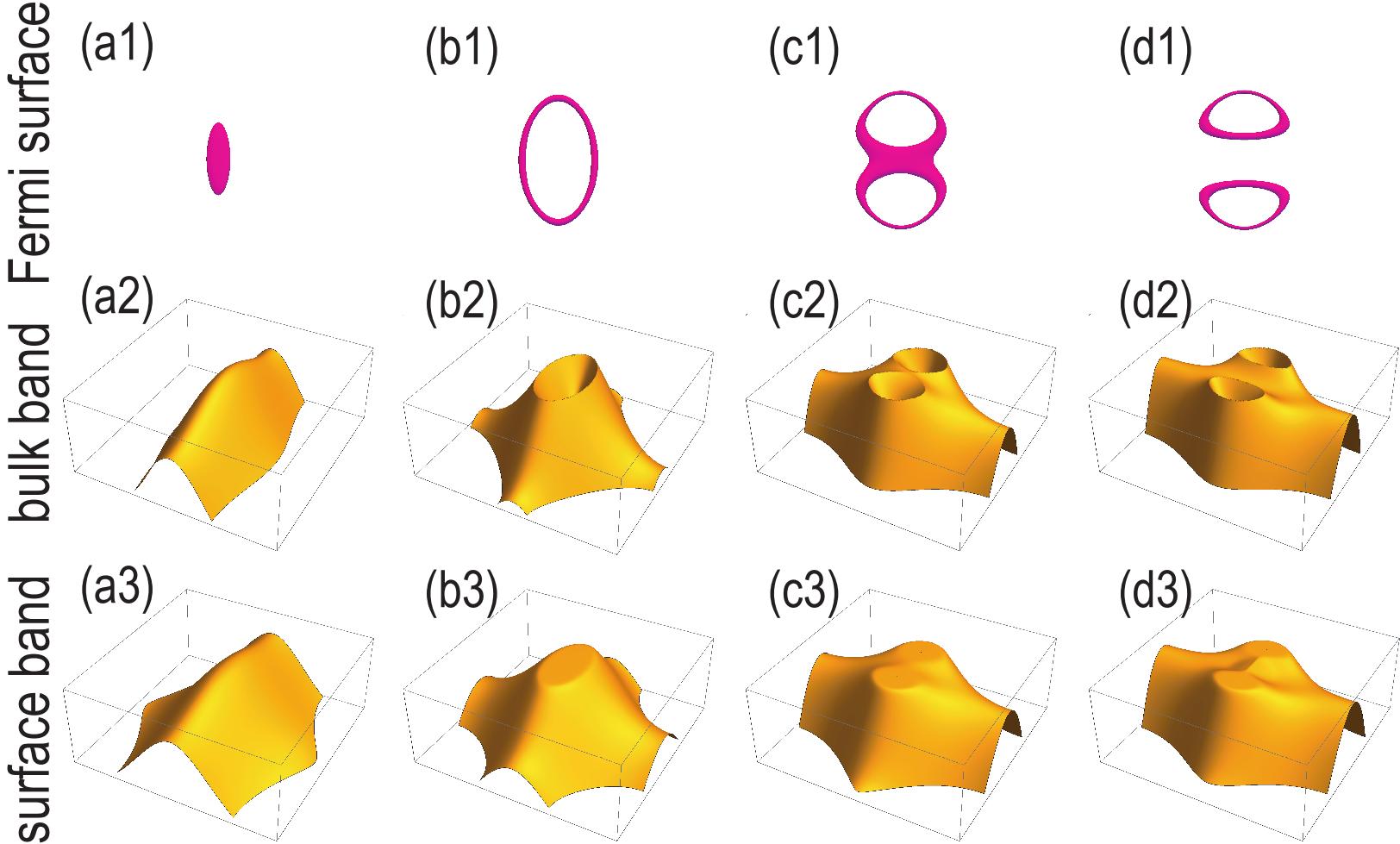}}
\caption{(a1)--(d1) Loop-nodal zero-energy Fermi surfaces for (a1) $%
t_{z}=t_{1}$, (b1) $t_{1}<t_{z}<t_{2}$, (c1) $t_{z}=t_{2}$ and (d1) $%
t_{1}<t_{z}<t_{2}$. (a2)--(d2) Band structures along $k_{x}=0$ plane.
(a3)--(d3) Drum-head surface states of the valence band along the $y$-$z$
plane. $t_{x}=t_{y}=1$, $\protect\lambda _{x}=\protect\lambda _{y}=1$; $m=-3$%
, $m_{\text{Loop}}=0.75$. In (a2)--(d3), only the valence bands are shown
for clarity.}
\label{FigLoop}
\end{figure}

\medskip\noindent\textbf{Loop-nodal semimetals.} 
We first study the loop nodal phase described by the Hamiltonian $H_{\text{Loop}}$. The energy
spectrum is simply given by 
\begin{equation}
E=\pm \sqrt{\lambda _{x}^{2}\sin ^{2}k_{x}+\left( \sqrt{\lambda _{y}^{2}\sin
^{2}k_{y}+M^{2}}\pm \left\vert m_{\text{Loop}}\right\vert \right) ^{2}}.
\end{equation}
The loop-nodal Fermi surface is obtained by solving $E\left( \mathbf{k}%
\right) =0$. It follows that $k_{x}=0$ and 
\begin{equation}
\lambda _{y}^{2}\sin ^{2}k_{y}+M^{2}\left( 0,k_{y},k_{z}\right) =m_{\text{%
Loop}}^{2}.  \label{LoopX}
\end{equation}
Loop nodes at zero energy exist in the $k_{x}=0$ plane. They are protected
by the mirror symmetry $M_{x}=\tau _{z}\mu _{z}\sigma _{x}$ with respect to
the $k_{x}=0$ plane and the $PT$ symmetry\cite{FangFu,BJYang}. We show the
band structure along the $k_{x}=0$ plane in Fig.\ref{FigLoop}(a2)--(d2). We
see clearly that the loop node structures are formed at the Fermi energy in
Fig.\ref{FigLoop}(b2)--(d2). These loop nodes are also observed as the
drum-head surface states, which are partial flat bands surrounded by the
loop nodes as shown in Fig.\ref{FigLoop}(b3)--(d3). The low energy $2\times
2 $ Hamiltonian is given by 
\begin{equation}
H=\left( \sqrt{\lambda _{y}^{2}\sin ^{2}k_{y}+M^{2}}\pm \left\vert m_{\text{%
Loop}}\right\vert \right) \sigma _{z}+\lambda _{x}\sin k_{x}\sigma _{x},
\end{equation}
where $\sigma $ is the Pauli matrix for the reduced two bands.

\begin{figure*}[t]
\centerline{\includegraphics[width=0.85\textwidth]{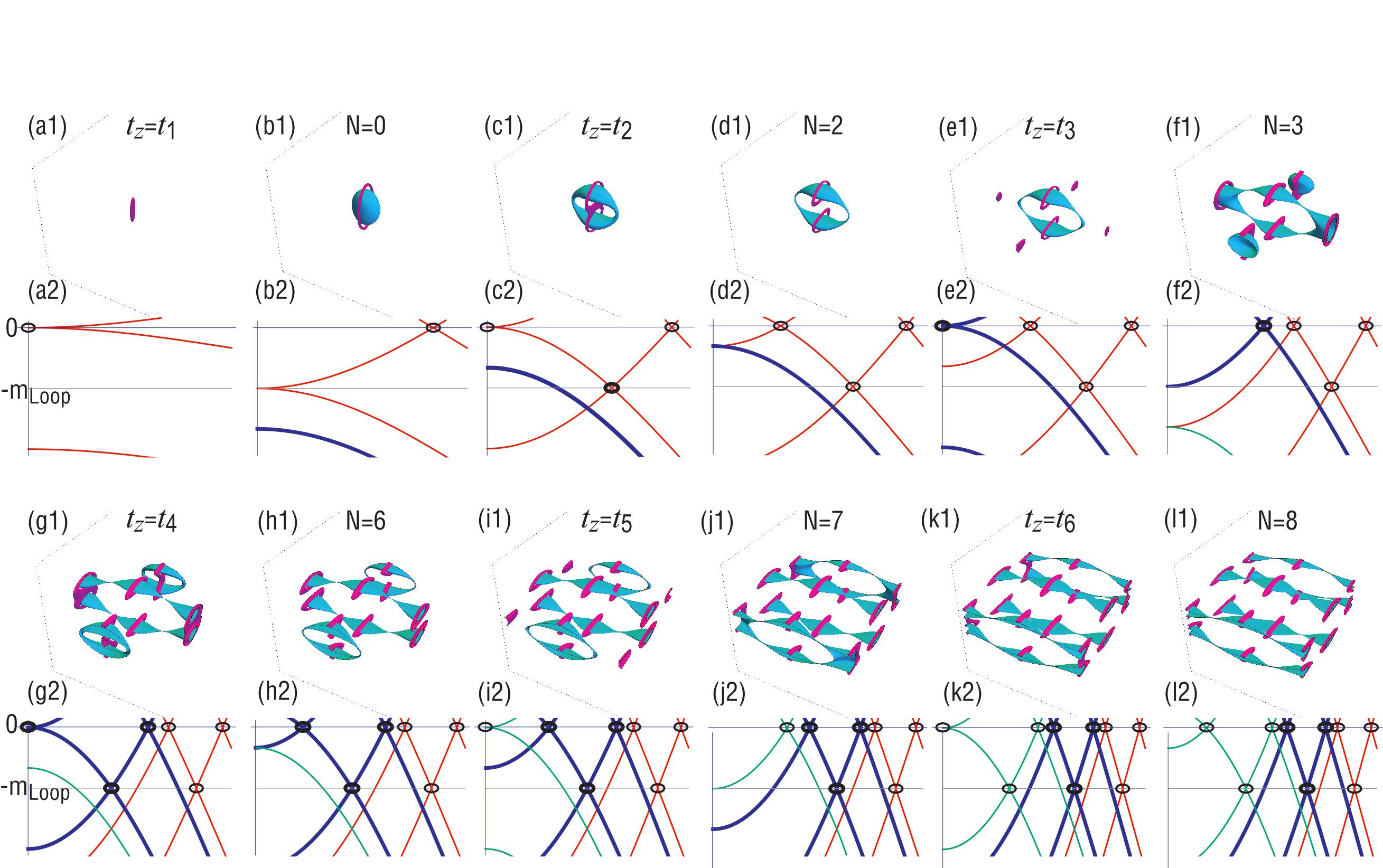}}
\caption{Evolution of linking structures for various $t_{z}$. (a) $t_z=t_1$,
(b) $t_1< t_z<t_2$, (c) $t_z=t_2$, (d) $t_2< t_z<t_3$, (e) $t_z=t_3$, (f) $%
t_3< t_z<t_4$, (g) $t_z=t_4$, (h) $t_4< t_z<t_5$, (i) $t_z=t_5$, (j) $t_5<
t_z<t_6$, (k) $t_z=t_6$ and (l) $t_z>t_6$. (a1)--(l1) Loop-nodal Fermi
surfaces at the zero-energy (magenta) and at $E=-m_{\text{Loop}} $ (cyan).
They are linked, whose linking number $N$ is shown in figures. (a2)--(l2)
Band structure along the $\Gamma $-$Z$ line (red), the $X$-$U$ and $Y$-$T$
lines (thick blue curves representing double degeneracy) and the $S$-$R$
line (green). Only the valence bands are shown for $0\leq k_z\leq \protect%
\pi $. Cross section of the loop nodes are marked in circles. The Parameters
are the same as in Fig.\protect\ref{FigLoop}.}
\label{FigLink}
\end{figure*}

In addition, there are loop nodes on the $k_{y}=0$ plane at $E=-m_{\text{Loop%
}}$, which are determined by 
\begin{equation}
\lambda _{x}^{2}\sin ^{2}k_{x}+\left( M\left( k_{x},0,k_{z}\right) -m_{\text{%
Loop}}\right) ^{2}=m_{\text{Loop}}^{2}.  \label{LoopY}
\end{equation}
We find the two loops determined by Eq.(\ref{LoopX}) and Eq.(\ref{LoopY})
are linked, as shown in Fig.\ref{FigLink}.

The system is a trivial insulator for $0\leq t_{z}<t_{1}$. One loop emerges
for $t_{1}<t_{z}<t_{2}$ [Fig.\ref{FigLoop}(b1)], which splits into two loops
for $t_{2}<t_{z}<t_{3}$, as shown in Fig.\ref{FigLoop}(d1). Correspondingly,
drum-head surface states, which are partial flat band within the loop nodes,
appear along the [100] surface [see Fig.\ref{FigLoop}(b3), (c3) and (d3)].

The emergence of the loop-nodal Fermi surface is understood in terms of the
band inversion\cite{BJYang,MoTe}, as shown in Fig.\ref{FigLink}. The number
of the loops are identical to the number of circles at the Fermi energy as
in Fig.\ref{FigLink}(a2)--(l2). When only one band is inverted along the $%
\Gamma $-$Z$ line, a single loop node appears [Fig.\ref{FigLink}(b1)]. When
two bands are inverted along the $\Gamma $-$Z$ line, two loop nodes appear
[Fig\ref{FigLink}(d1)]. In the similar way, additional loops appear when
additional bands are inverted along the $X$-$U$ and $Y$-$T$ lines [Fig.\ref%
{FigLink}(f1)], and it is split into two loops [Fig.\ref{FigLink}(h1)] as $%
t_{z}$ increases. In the final process, a loop appears along the $S$-$R$
line [Fig.\ref{FigLink}(j1)], which splits into two loops [Fig.\ref{FigLink}%
(l1)].

It has been argued\cite{BJYang,MoTe} that a new topological nature of
loop-nodal semimetals becomes manifest when we plot the loop-nodal Fermi
surfaces at the band crossing energies, where one is at the Fermi energy and
the other is at $E=-m_{\text{Loop}}$ in the occupied band. We show them in
Fig.\ref{FigLink}. Along the $\Gamma $-$Z$ line, the other band crossing
occurs at $\pm m_{\text{Loop}}$ with 
\begin{equation}
\left\vert k_{z}\right\vert =\arccos \left[ \left( m_{\text{Loop}%
}-m-2t\right) /t_{z}\right] .
\end{equation}
Along the $X$-$U$ and $Y$-$T$ lines, the band crossing occurs also at $\pm
m_{\text{Loop}}$ with 
\begin{equation}
\left\vert k_{z}\right\vert =\arccos \left[ -m/t_{z}\right] .
\end{equation}
Along the $S$-$R$ line, the band crossing occurs also at $\pm m_{\text{Loop}%
} $ with 
\begin{equation}
\left\vert k_{z}\right\vert =\arccos \left[ \left( -m+2t\right) /t_{z}\right]%
.
\end{equation}
As a result, it is enough to plot the Fermi surfaces at $E=0$ and $E=-m_{%
\text{Loop}}$. The linking number $N$ increases as $t_{z}$ increases, where
even the linking number $N=8$ is realized as in Fig.\ref{FigLink}(l1).

\medskip\noindent\textbf{2D TI, TCI and SOTI.} 
At this stage it is convenient to study the 2D models by setting $t_{z}=\lambda _{z}=0$. It
follows from (\ref{3DTopBound}) that the 2D topological phase boundaries are
given by 
\begin{equation}
\left( m+\eta _{x}t_{x}+\eta _{y}t_{y}\right) ^{2}=m_{\text{Loop}}^{2}+m_{%
\text{SOTSM}}^{2},  \label{2DTopBound}
\end{equation}
where $\eta _{x}=\pm 1$ and $\eta _{y}=\pm 1$. Depending on the way to
introduce the mass parameters there are three phases, i.e., TIs, TCIs and
SOTIs.

\begin{figure}[t]
\centerline{\includegraphics[width=0.45\textwidth]{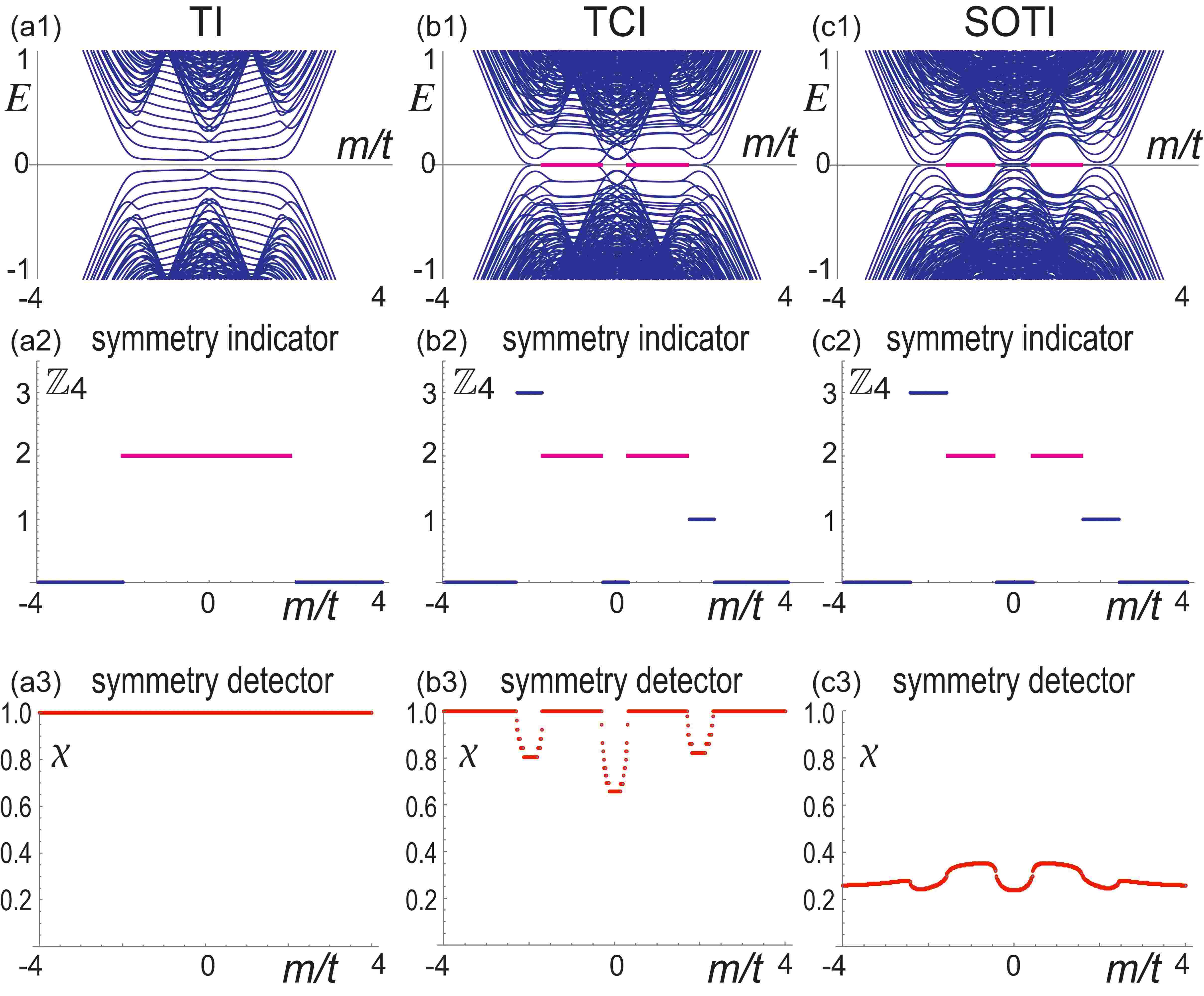}}
\caption{(a1)-(c1) Energy spectrum as a function of $m/t$ for TI, TCI and
SOTI phases. (a2)-(c2) Corresponding $\mathbb{Z}_{4}$ index. (a3)-(c3)
Corresponding mirror-symmetry detector $\protect\chi$. It follows that $\protect\chi =1$ for the TI
and the insulating phase of the TCI, and that $\protect\chi \neq 1$ for the
SOTI since the mirror symmetry is broken.}
\label{FigZ4}
\end{figure}

The topological number is known to be the $\mathbb{Z}_{4}$ index protected
by the inversion symmetry in three dimensions\cite{Po,Khalaf,Song2,MoTe}.
This is also the case in two dimensions. It is defined by 
\begin{equation}
\kappa _{1}\equiv \frac{1}{4}\sum_{K\in \text{TRIMs}}\left(
n_{K}^{+}-n_{K}^{-}\right) ,
\end{equation}%
where $n_{K}^{\pm }$ is the number of occupied band with the parity $\pm $.
There is a relation\cite{Po,Khalaf,Song2} 
\begin{equation}
\text{mod}_{2}\kappa _{1}=\nu,
\end{equation}%
where $\nu $ is the $\mathbb{Z}_{2}$ index characterizing the
time-reversal invariant TIs. We find from Fig.\ref{FigZ4}(c1) that $\kappa _{1}=0,2$ in the TI phase, 
which implies that it is trivial in the viewpoint of the time-reversal invariant topological insulators.

\begin{figure}[t]
\centerline{\includegraphics[width=0.45\textwidth]{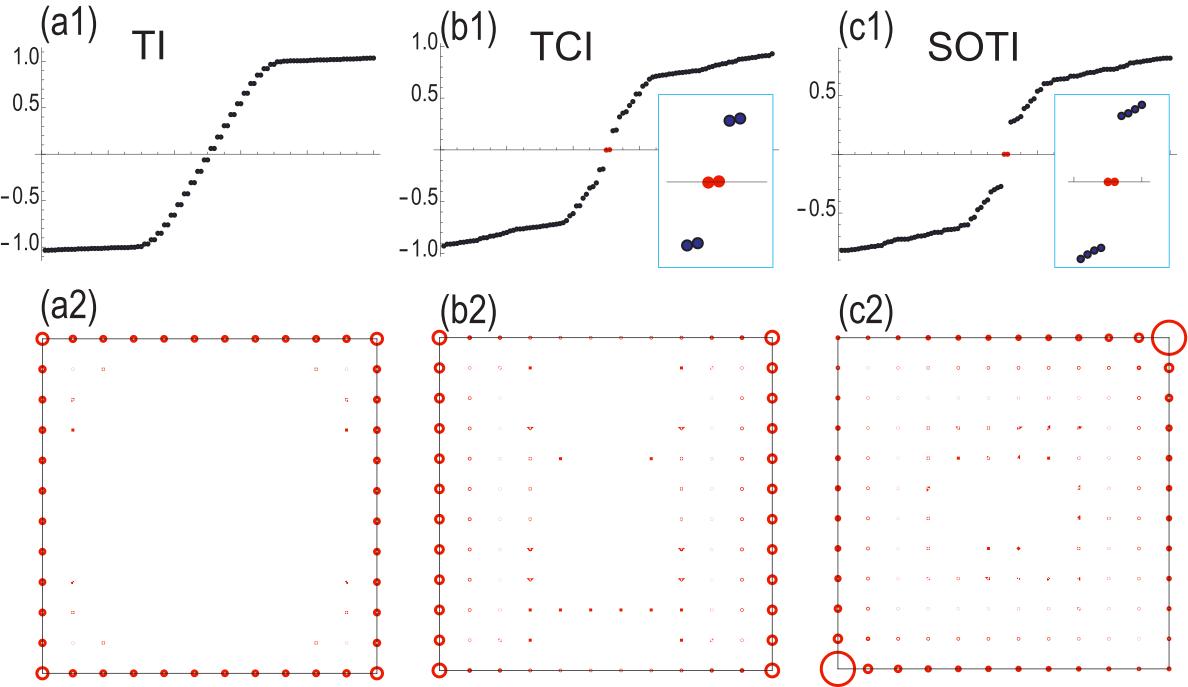}}
\caption{(a1)--(c1) Eigenvalues of the sample in a square geometry, where
the insets show the zero-energy states in red. The vertical axis is the
energy. (a2)--(c2) corresponding LDOS of the zero-energy states. The
amplitude is represented by the radius of the circles. We have set $%
t_{x}=t_{y}=m=\protect\lambda =1$ and $m_{\text{Loop}}=m_{\text{SOTSM}}=0.3$%
. }
\label{FigSquare}
\end{figure}

We show the LDOS for TI, TCI and SOTI in Fig.\ref{FigSquare}. (i) When $m_{%
\text{Loop}}=m_{\text{SOTSM}}=0$ and $\left\vert m\right\vert <2t$, the
system is a TI with $\kappa _{1}=2$, where topological edge states appear
for all edges [See Fig.\ref{FigSquare}(a)]. We show the energy spectrum and
the $Z_{4}$ index in Fig.\ref{FigZ4}(a1) and (a2), respectively. The energy
spectrum is two-fold degenerate since there is the symmetry $P\bar{T}=\mu
_{y}$ such that $\left( P\bar{T}\right) ^{-1}H_{0}(k)P\bar{T}=H_{0}(k)$.
Furthermore, there is the mirror symmetry $M_{x}=i\tau _{z}\mu _{z}$ such
that $M_{x}^{-1}H_{\text{Loop}}\left( k_{x},k_{y}\right) M_{x}=H_{\text{Loop}%
}\left( -k_{x},k_{y}\right) $. (ii) When $m_{\text{Loop}}\neq 0$ and $m_{%
\text{SOTSM}}=0$, the system is a TCI, where topological edge states appear
only for two edges [See Fig.\ref{FigSquare}(b)]. The energy spectrum and the 
$Z_{4}$ index are shown in Fig.\ref{FigZ4}(b1) and (b2). The symmetry $P\bar{%
T}$\ is broken for $m_{\text{Loop}}\neq 0$ and the two-fold degeneracy is
resolved. On the other hand, the mirror symmetry $M_{x}$ remains preserved.
(iii) Finally, when $m_{\text{Loop}}\neq 0$ and $m_{\text{SOTSM}}\neq 0$,
the system is a SOTI, where two corner states emerge [See Fig.\ref{FigSquare}%
(c)]. The energy spectrum and the $Z_{4}$ index are shown in Fig.\ref{FigZ4}%
(c1) and (c2). The mirror symmetry is broken in the SOTI phase. In TCI and
SOTI phases, there are regions where $\kappa _{1}=1$, $3$. However, in this
region, the system is semimetallic and the $\kappa _{1}$ index has no
meaning. 

The $Z_{4}$ index takes the same value for the TI, TCI and SOTI phases, and
hence it cannot differentiate them. Indeed, because there is no band gap
closing between them\cite{GapClose}, the symmetry indicator cannot change
its value\cite{Khalaf}. A natural question is whether there is another
topological index to differentiate them. We propose the symmetry detector
discriminating whether the symmetry is present or not.

The TI and TCI are differentiated whether the symmetry $P\bar{T}$\ is
present or not. The band is two-fold degenerate due to the symmetry $P\bar{T}
$\ in the TI phase, where we can define a topological index by%
\begin{equation}
\zeta =\text{Mod}_{4}\sum_{K\in \text{TRIMs}}\frac{\text{Pf}\left[ w\right] 
}{\sqrt{\det \left[ w\right] }}
\end{equation}%
with%
\begin{equation}
w_{ij}=\left\langle \psi _{i}\left( -K\right) \right\vert P\bar{T}\left\vert
\psi _{j}\left( K\right) \right\rangle .
\end{equation}%
where $i$ and $j$ are the two-fold degenerated band index. It is only
defined for the TI phase, where it gives the same result as $\kappa _{1}$.\ 
On the other hand, it is ill-defined for the TCI and SOTI phases
since there is no band degeneracy.

The TCI and SOTI are differentiated by the mirror-symmetry detector defined
by%
\begin{equation}
\chi \equiv \chi _{0}^{+}\chi _{\pi }^{+}\chi _{0}^{-}\chi _{\pi }^{-},
\end{equation}%
where 
\begin{equation}
\chi _{\alpha }^{\pm }\equiv \frac{-i}{2\pi }\left. \int_{0}^{2\pi
}\left\langle \psi \right\vert M_{x}\left\vert \psi \right\rangle
dk_{y}\right\vert _{k_{x}=\alpha }
\end{equation}%
is the mirror symmetry indicator\cite{Switch} along the axis $k_{x}=\alpha $
with $\alpha =0,\pi $, and $\pm $ indicates the band index under the Fermi
energy. It is $\chi =1$ when there is the mirror symmetry. On the other
hand, it is $\chi \neq 1$ when there is no mirror symmetry since $\left\vert
\psi \right\rangle $ is not the eigenstate of the mirror operator. In
addition, it is $\chi \neq 1$\ when the system is metallic since $%
\left\langle \psi \right\vert M_{x}\left\vert \psi \right\rangle $ changes
its value at band gap closing points. See Figs.\ref{FigZ4}(a3)-(c3).\ In Fig.%
\ref{FigZ4}(a3), we find always $\chi =1$\ since the mirror symmetry is
preserved, where we cannot differentiate the topological and trivial
phases. On the other hand, in Fig.\ref{FigZ4}(b3), there are regions with $%
\chi \neq 1$ where the system is metallic. Finally, we find $\chi \neq 1$ in
Fig.\ref{FigZ4}(c3)\ since the mirror symmetry is broken.

\medskip\noindent\textbf{SOTSM.} 
A 3D SOTSM is constructed by considering $%
k_{z}$ dependent mass term in the 2D SOTI model\cite{Lin,EzawaKagome,MagHOTI}%
. We set $t_{z}\neq 0$, while keeping $\lambda _{z}=0$ in the 2D SOTI model.
The properties of the SOTSM are derived by the sliced Hamiltonian $H(k_{z})$
along the $k_{z}$ axis, which gives a 2D SOTI model with $k_{z}$ dependent
mass term $M(k_{z})$. The bulk band gap closes at 
\begin{equation}
M^{2}\left( k_{z}\right) =m_{\text{Loop}}^{2}+m_{\text{SOTSM}}^{2}.
\end{equation}
On the other hand, there emerge hinge-arc states connecting the two gap
closing points. Accordingly, the topological corner states in the 2D\ SOTI
model evolves into hinge-states, whose dispersion forms flat bands as shown
in Fig.\ref{FigCube}(c4).

\medskip\noindent\textbf{Magnetic control of hinges in SOTI.} 
Hinge states are analogous to edge states in two-dimensional topological insulators.
Without applying external field, spin currents flow. On the other hand,
once electric field is applied, charge current carrying a quantized
conductance flows. We show that the current is controlled by the direction
of magnetization as in the case of topological edge states.

With the inclusion of the $H_{\text{SO}}$, the system turns into a SOTI,
which has topological hinge states. We study the effects of the Zeeman term,
where the Hamiltonian is described by $H_{\text{SOTI}}$ together with the
Zeeman term 
\begin{equation}
H_{\text{Z}}=B_{x}\sigma _{x}+B_{y}\sigma _{y}+B_{z}\sigma _{z},
\end{equation}
which will be introduced by magnetic impurities, magnetic proximity effects
or applying magnetic field.

\begin{figure}[t]
\centerline{\includegraphics[width=0.45\textwidth]{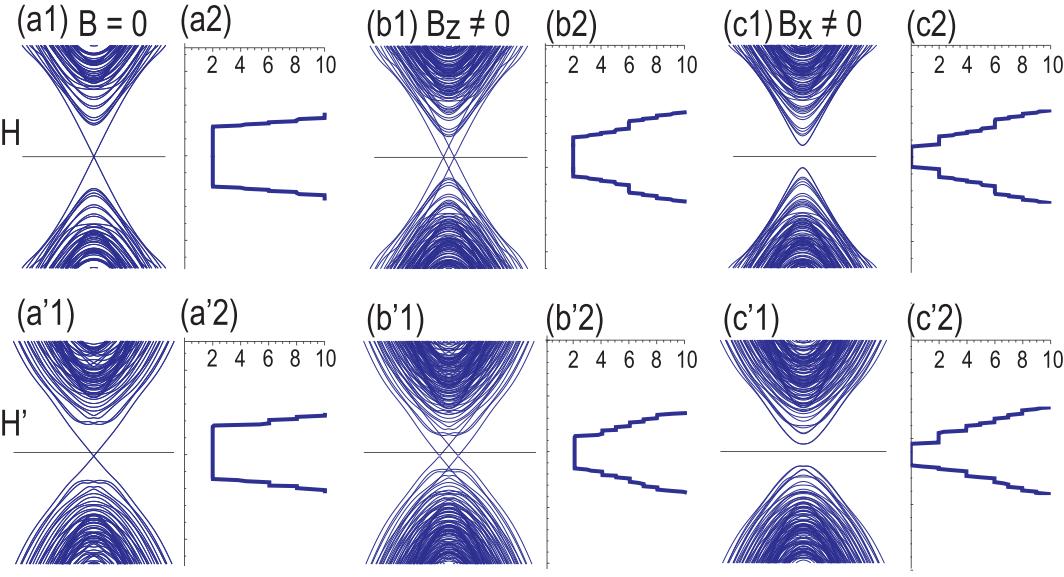}}
\caption{Band structures for hinge states (a1) without magnetic field, (b1)
with magnetic field along the $z$ direction and (c1) with magnetic field
along the $x$ direction for the chiral-symmetric Hamiltonian $H_{\text{SOTI}%
} $. Hinge states are depicted in red. (a2)--(c2) Corresponding ones for the
chiral-nonsymmetric Hamiltonian $H_{\text{SOTI}}^{\prime }$. (a2)--(c2) and
(a'2)--(c'2) The conductance is quantized proportional to the number of
bands in various cases. }
\label{FigBHinge}
\end{figure}

We show the hinge states in the absence and the presence of magnetization in
Fig.\ref{FigBHinge}. Helical hinge states appear in its absence [see Fig.\ref%
{FigBHinge}(a1)]. They are shifted in the presence of the $B_{z}$ term [see
Fig.\ref{FigBHinge}(b1)]. On the other hand, they are gapped out when the $%
B_{x}$ or $B_{y}$ term exists [see Fig.\ref{FigBHinge}(c1)].

For comparison, we also show the hinge states calculated from the
chiral-nonsymmetric Hamiltonian $H^{\prime }_{\text{SOTI}}$ [see Fig.\ref%
{FigBHinge}(a2)--(c2)]. The band structure is almost symmetric with respect
to the Fermi energy.

By taking into the fact that the $\sigma _{z}$ is a good quantum number, the
low energy theory of the hinge states is well described by 
\begin{equation}
H=\hbar v_{\text{F}}k_{z}\sigma _{z}.
\end{equation}
In the presence of the external magnetic field, it is modified as 
\begin{equation}
H=\hbar v_{\text{F}}k_{z}\sigma _{z}+B_{x}\sigma _{x}+B_{y}\sigma
_{y}+B_{z}\sigma _{z},
\end{equation}
which is easily diagonalized to be 
\begin{equation}
E=\pm \sqrt{(\hbar v_{\text{F}}k_{z}+B_{z})^{2}+B_{x}^{2}+B_{y}^{2}}.
\end{equation}
It well reproduces the results based on the tight binding model shown in Fig.%
\ref{FigBHinge}.

One of the intrinsic features of a topological hinge state is that it
conveys a quantized conductance in the unit of $e^{2}/h$. We have calculated
the conductance of the hinge states in Fig.\ref{FigBHinge} based on the
Landauer formalism\cite{Datta,Rojas,Nikolic,Li,EzawaAPL}. In terms of
single-particle Green's functions, the conductance $\sigma (E)$ at the
energy $E$ is given by\cite{Datta} 
\begin{equation}
\sigma (E)=(e^{2}/h)\text{Tr}[\Gamma _{\text{L}}(E)G_{\text{D}}^{\dag
}(E)\Gamma _{\text{R}}(E)G_{\text{D}}(E)],  \label{G}
\end{equation}
where $\Gamma _{\text{R(L)}}(E)=i[\Sigma _{\text{R(L)}}(E)-\Sigma _{\text{%
R(L)}}^{\dag }(E)]$ with the self-energies $\Sigma _{\text{L}}(E)$ and $%
\Sigma _{\text{R}}(E)$, and 
\begin{equation}
G_{\text{D}}(E)=[E-H_{\text{D}}-\Sigma _{\text{L}}(E)-\Sigma _{\text{R}%
}(E)]^{-1},  \label{StepA}
\end{equation}
with the Hamiltonian $H_{\text{D}}$ for the device region. The self energies 
$\Sigma _{\text{L}}(E)$ and $\Sigma _{\text{R}}(E)$ are numerically obtained
by using the recursive method\cite{Datta,Rojas,Nikolic,Li,EzawaAPL}.

The conductance is quantized, which is proportional to the number of bands.
When there is no magnetization or the magnetization is along the $z$ axis,
the conductance is $2$ since there are two topological hinges. On the other
hand, once there is in-plane magnetization, the conductance is switched off
since the hinge states are gapped. It is a giant magnetic resistor\cite%
{Rachel}, where the conductance is controlled by the magnetization direction.

\Section{Conclusion}

We have studied chiral-symmetric models to describe SOTIs and loop-nodal
semimetals in transition metal dichalcogenides. The Hamiltonian is
analytically diagonalized due to the chiral symmetry. We have obtained
analytic formulas for various phases including loop-nodal semimetals, 2D
SOTIs, 3D SOTSMs and 3D SOTIs. We have proposed the symmetry detector 
discriminating whether the symmetry is present or not. It can
differentiate topological phases to which the symmetry indicator yields an identical value. 
Furthermore, we have proposed a
topological device, where the conductance is switched by the direction of
magnetization. Our results will open a way to topological devices based on
transition metal dichalcogenides.

\Section{Acknowledgements}

The author is very much grateful to N. Nagaosa for helpful discussions on
the subject. This work is supported by the Grants-in-Aid for Scientific
Research from MEXT KAKENHI (Grant Nos.JP17K05490, JP15H05854 and No.
JP18H03676). This work is also supported by CREST, JST (JPMJCR16F1).

\Section{Author contributions} M.E. conceived the idea, performed the
analysis, and wrote the manuscript.

\Section{Additional information} \textbf{Competing financial and
non-financial interests:} The author declares no competing financial and
non-financial interests.


\begin{thebibliography}{99}
\bibitem{Fan} Zhang, F., Kane, C. L. \& Mele, E.J. Surface State
Magnetization and Chiral Edge States on Topological Insulators. \textit{%
Phys. Rev. Lett.} \textbf{110}, 046404 (2013).

\bibitem{Science} Benalcazar, W. A., Bernevig, B. A. \& Hughes, T. L.
Quantized electric multipole insulators. \textit{science} \textbf{357}, 61
(2017). 

\bibitem{APS} Schindler, F., Cook, A., Vergniory, M. G. \& Neupert, T.
Higher-order Topological Insulators and Superconductors. \textit{in APS
March Meeting} (2017).

\bibitem{Peng} Peng, Y., Bao, Y. \& von Oppen, F. Boundary Green functions
of topological insulators and superconductors. \textit{Phys. Rev. B} \textbf{%
95}, 235143 (2017).

\bibitem{Lang} Langbehn, J., Peng, Y., Trifunovic, L., von Oppen, F. \&
Brouwer, P. W. Reflection-Symmetric Second-Order Topological Insulators and
Superconductors. \textit{Phys. Rev. Lett.} \textbf{119}, 246401 (2017).

\bibitem{Song} Song, Z., Fang, Z. \& Fang, C. $(d-2)$-Dimensional Edge
States of Rotation Symmetry Protected Topological States. \textit{Phys. Rev.
Lett.} \textbf{119}, 246402 (2017).

\bibitem{Bena} Benalcazar, W. A., Bernevig, B. A. \& Hughes, T. L. Electric
multipole moments, topological multipole moment pumping, and chiral hinge
states in crystalline insulators. \textit{Phys. Rev.} \textbf{B} \textbf{96}%
, 245115 (2017).

\bibitem{Schin} Schindler, F., Cook, A. M., Vergniory, M. G., Wang, Z.,
Parkin, S. S. P., Bernevig, B. A., \& Neupert, T. Higher-order topological
insulators. \textit{Science Advances} \textbf{4}, eaat0346 (2018).

\bibitem{FuRot} Fang, C., Fu, L. Rotation Anomaly and Topological
Crystalline Insulators. arXiv:1709.01929.

\bibitem{EzawaKagome} Ezawa, M. Higher-Order Topological Insulators and
Semimetals on the Breathing Kagome and Pyrochlore Lattices. \textit{Phys.
Rev. Lett.} \textbf{120}, 026801 (2018).

\bibitem{Gei} Geier, M., Trifunovic, L., Hoskam, M., \& Brouwer, P. W.
Second-order topological insulators and superconductors with an order-two
crystalline symmetry. \textit{Phys. Rev. B} \textbf{97}, 205135 (2018).

\bibitem{Lin} Lin, M. \& Hughes, T. L. Topological Quadrupolar Semimetals. 
\textit{Phys. Rev. B} \textbf{98}, 241103 (2018).

\bibitem{MagHOTI} Ezawa, M. Magnetic second-order topological insulators and
semimetals. \textit{Phys. Rev. B} \textbf{97}, 155305 (2018).

\bibitem{Kha} Khalaf, E. Higher-order topological insulators and
superconductors protected by inversion symmetry. \textit{Phys. Rev. B} 
\textbf{97}, 205136 (2018).

\bibitem{HexaHOTI} Ezawa, M. Strong and weak second-order topological
insulators with hexagonal symmetry and $Z_3$ index. \textit{Phys. Rev. B} 
\textbf{97}, 241402(R) (2018).

\bibitem{EzawaPhos} Ezawa, M. Minimal models for Wannier-type higher-order
topological insulators and phosphorene. \textit{Phys. Rev. B} \textbf{98},
045125 (2018).

\bibitem{Bis} Schindler, F., Wang, Z., Vergniory, M. G., Cook, A. M.,
Murani, A., Sengupta, S., Kasumov, A. Y., Deblock, R., Jeon, S., Drozdov,
I., Bouchiat, H., Gueron, S., Yazdani, A., Bernevig, B. A., \& Neupert, T.
Higher-order topology in bismuth. \textit{Nature Physics} \textbf{14}, 918
(2018).

\bibitem{TQP} Bradlyn B., Elcoro, L., Cano, J., Vergniory, M. G., Wang, Z.,
Felser, C., Aroyo, M. I. \& Bernevig, B. A. Topological quantum chemistry. 
\textit{Nature} \textbf{547}, 298 (2017).

\bibitem{MoTe2} Tang, F., Po, H. C., Vishwanath, A. \& Wan, X. Efficient
Topological Materials Discovery Using Symmetry Indicators. arXiv:1805.07314.

\bibitem{MoTe} Wang, Z., Wieder, B. J., Li, J., Yan, B. \& Bernevig, B. A.
Higher-Order Topology, Monopole Nodal Lines, and the Origin of Large Fermi
Arcs in Transition Metal Dichalcogenides XTe2 (X=Mo,W). arXiv:1806.11116.

\bibitem{FangLoop} Fang, C., Chen, Y., Kee, H.-Y. \& Fu L. Topological nodal
line semimetals with and without spin-orbital coupling. \textit{Phys. Rev. B}
\textbf{92}, 081201(R) (2015).

\bibitem{Kim} Kim, Y., Wieder, B. J., Kane, C. L. \& Rappe, A. M. Dirac Line
Nodes in Inversion-Symmetric Crystals. \textit{Phys. Rev. Lett.} \textbf{115}%
, 036806 (2015).

\bibitem{Yu} Yu, R., Weng, H., Fang, Z., Dai, X. \& Hu, X. Topological
Node-Line Semimetal and Dirac Semimetal State in Antiperovskite Cu$_3$PdN. 
\textit{Phys. Rev. Lett.} \textbf{115}, 036807 (2015).

\bibitem{Chan} Chan, Y.-H., Chiu, C.-K., Chou, Y. \& Schnyder, A. P. Ca$_3$P$%
_2$ and other topological semimetals with line nodes and drumhead surface
states. \textit{Phys. Rev. B} \textbf{93}, 205132 (2016).

\bibitem{CFang} Song, Z., Zhang, T. \& Fang, C. Diagnosis for Nonmagnetic
Topological Semimetals in the Absence of Spin-Orbital Coupling \textit{Phys.
Rev. X} \textbf{8}, 031069 (2018).

\bibitem{WChen} Chen, W., Lu, H.-Z. \& Hou, J.-M. Topological semimetals
with a double-helix nodal link. \textit{Phys. Rev. B} \textbf{96}, 041102
(2017).

\bibitem{ZYan} Yan, Z., Bi, R., Shen, H., Lu, L., Zhang, S.-C. \& Wang, Z.
Nodal-link semimetals. \textit{Phys. Rev. B} \textbf{96}, 041103(R) (2017).

\bibitem{PYChang} Chang, P.-Y. \& Yee, C.-H. Weyl-link semimetals. \textit{%
Phys. Rev. B} \textbf{96}, 081114 (2017).

\bibitem{EzawaHopf} Ezawa, M. Topological semimetals carrying arbitrary Hopf
numbers: Fermi surface topologies of a Hopf link, Solomon's knot, trefoil
knot, and other linked nodal varieties. \textit{Phys. Rev. B} \textbf{96},
041202(R) (2017).

\bibitem{HasanPRL} Chang, G., Xu, S.-Y., Zhou, X., Huang, S.-M., Singh, B.,
Wang, B., Belopolski, I., Yin, J., Zhang, S., Bansil, A., Lin, H., Hasan, M.
Z. Topological Hopf and Chain Link Semimetal States and Their Application to
Co$_2$MnGa. \textit{Phys. Rev. Lett.} \textbf{119}, 156401 (2017).

\bibitem{BJYang} Ahn, J., Kim, Y. \& Yang, B.-J. Band Topology and Linking
Structure of Nodal Line Semimetals with Z$_2$ Monopole Charges. \textit{%
Phys. Rev. Lett.} \textbf{121}, 106403 (2018).

\bibitem{MagTI1} Chen, Y. L., Chu, J.-H., Analytis, J. G., Liu, Z. K.,
Igarashi, K., Kuo, H.-H., Qi, X. L., Mo, S. K., Moore, R. G., Lu, D. H.,
Hashimoto, Sasagawa, T., Zhang, S. C., Fisher, I. R., Hussain, Z. \& Shen,
Z. X. Massive Dirac Fermion on the Surface of a Magnetically Doped
Topological Insulator. \textit{Science} \textbf{329}, 659 (2010).

\bibitem{MagTI2} J. Zhang, Chang, C.-Z., Tang P., Zhang, Z., Feng, X., Li
,K., Wang, L.-l. , Chen, X., Liu, C., Duan, W., He, K., Xue, Q.-K. , Ma, M.
\& WangY., Topology-Driven Magnetic Quantum Phase Transition in Topological
Insulators. \textit{Science} \textbf{339}, 1582 (2013).

\bibitem{MagTI3} Chang, C.-Z., et.al., Experimental Observation of the
Quantum Anomalous Hall Effect in a Magnetic Topological Insulator. Science
340, 167 (2013).

\bibitem{MagTI4} Checkelsky, J. G., Yoshimi, R., Tsukazaki, A., Takahashi,
K. S., Kozuka, Y., Falson, J. , Kawasaki, M. \& Tokura, Y. Trajectory of the
anomalous Hall effect towards the quantized state in a ferromagnetic
topological insulator. Nat. Phys. 10, 731 (2014).

\bibitem{Rachel} Rachel, S. \& Ezawa, M. Giant magnetoresistance and perfect
spin filter in silicene, germanene, and stanene. \textit{Phys. Rev. B}bf{89}%
, 195303 (2014).

\bibitem{Switch} Ezawa, M. Topological Switch between Second-Order
Topological Insulators and Topological Crystalline Insulators. \textit{Phys.
Rev. Lett.} \textbf{121}, 116801 (2018).

\bibitem{KaneMele} Kane, C. L. \& Mele, E. J., $Z_2$ Topological Order and
the Quantum Spin Hall Effect. \textit{Phys. Rev. Lett.} \textbf{95}, 146802
(2005): Quantum Spin Hall Effect in Graphene. \textit{Phys. Rev. Lett.} 
\textbf{95}, 226801 (2005).

\bibitem{EzawaReview} Ezawa, M. Monolayer Topological Insulators: Silicene,
Germanene, and Stanene. \textit{J. Phys. Soc. Jpn.} 84, 121003 (2015).

\bibitem{GapClose} Ezawa, M. Tanaka, Y and Nagaosa, N., Topological Phase
Transition without Gap Closing \textit{Scientific Reports} 3, 2790 (2013)

\bibitem{Datta} Datta, S. \textit{Electronic Transport in Mesoscopic Systems}
(Cambridge University Press, Cambridge, England, 1995): \textit{Quantum
transport: atom to transistor} (Cambridge University Press, England, 2005).

\bibitem{Rojas} Mu\~{n}oz-Rojas, F., Jacob, D., Fern\'{a}ndez-Rossier, J. \&
Palacios, J. J. Coherent transport in graphene nanoconstrictions. \textit{%
Phys. Rev. B} \textbf{74}, 195417 (2006).

\bibitem{Nikolic} Z\^{a}rbo, L. P. \& Nikoli\'{c}, B. K. Condensed Matter:
Electronic Structure, Electrical, Magnetic and Optical Properties. \textit{%
EPL}, \textbf{80} 47001 (2007): Areshkin, D. A. \& Nikoli\'{c}, B. K. I-V
curve signatures of nonequilibrium-driven band gap collapse in magnetically
ordered zigzag graphene nanoribbon two-terminal devices. \textit{Phys. Rev. B%
} \textbf{79}, 205430 (2009).

\bibitem{Li} Li, T. C. \& Lu, S.-P. Quantum conductance of graphene
nanoribbons with edge defects. \textit{Phys. Rev. B} \textbf{77}, 085408
(2008).

\bibitem{EzawaAPL} Ezawa, M. Quantized conductance and field-effect
topological quantum transistor in silicene nanoribbons. \textit{Appl. Phys.
Lett.} \textbf{102}, 172103 (2013).

\bibitem{Po} Po, H. C., Vishwanath, A. \& Watanabe, H. Symmetry-based
indicators of band topology in the 230 space groups. \textit{Nat. Comm.} 
\textbf{8}, 50 (2017).

\bibitem{Song2} Song, Z., Zhang, T., Fang, Z. \& Fang, C. Quantitative
mappings between symmetry and topology in solids. \textit{Nat. Com.} \textbf{%
9}, 3530 (2018).

\bibitem{Khalaf} Khalaf, E., Po, H. C., Vishwanath, A. \& Watanabe, H.
Symmetry Indicators and Anomalous Surface States of Topological Crystalline
Insulators. \textit{Phys. Rev. X} \textbf{8}, 031070 (2018).

\bibitem{FangFu} Fang, C., Chen, Y., Kee, H.-Y. \& Fu, L. Topological nodal
line semimetals with and without spin-orbital coupling. \textit{Phys. Rev. B}
\textbf{92}, 081201(R) (2015).
\end{thebibliography}
\end{document}